\documentclass[aps,prl,showpacs,twocolumn,superscriptaddress]{revtex4}
\usepackage{bm,color}
\usepackage{graphicx}
\usepackage{amsmath}
\usepackage{color}

\begin{document}
\title {Spin Model for Nontrivial Magnetic Orders in the Inverse-Perovskite Antiferromagnets}

\author{Masahito Mochizuki}
\affiliation{Department of Physics and Mathematics, Aoyama Gakuin University, Sagamihara, Kanagawa 229-8558, Japan}
\affiliation{Department of Applied Physics, Waseda University, Okubo, Shinjuku-ku, Tokyo 169-8555, Japan}
\affiliation{PRESTO, Japan Science and Technology Agency, Kawaguchi, Saitama 332-0012, Japan}
\author{Masaya Kobayashi}
\affiliation{Department of Physics and Mathematics, Aoyama Gakuin University, Sagamihara, Kanagawa 229-8558, Japan}
\author{Reoya Okabe}
\affiliation{Department of Physics and Mathematics, Aoyama Gakuin University, Sagamihara, Kanagawa 229-8558, Japan}
\author{Daisuke Yamamoto}
\affiliation{Department of Physics and Mathematics, Aoyama Gakuin University, Sagamihara, Kanagawa 229-8558, Japan}
\begin{abstract}
Nontrivial magnetic orders in the inverse-perovskite manganese nitrides are theoretically studied by constructing a classical spin model describing the magnetic anisotropy and frustrated exchange interactions inherent in specific crystal and electronic structures of these materials. With a replica-exchange Monte-Carlo technique, a theoretical analysis of this model reproduces the experimentally observed triangular $\Gamma^{5g}$ and $\Gamma^{4g}$ spin ordered patterns and the systematic evolution of magnetic orders. Our work solves a 40-year-old problem of nontrivial magnetism for the inverse-perovskite manganese nitrides and provides a firm basis for clarifying the magnetism-driven negative thermal expansion phenomenon discovered in this class of materials.
\end{abstract}
\maketitle

\begin{figure}[b]
\includegraphics[scale=1.0]{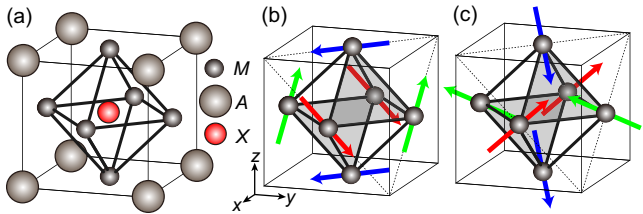}
\caption{(color online). (a) Inverse-perovskite structure. (b) Triangular $\Gamma^{5g}$ spin order observed in Mn$_3$ZnN and Mn$_3$GaN. (c) Triangular $\Gamma^{4g}$ spin order observed in Mn$_3$NiN and Mn$_3$AgN.}
\label{Fig1}
\end{figure}
Noncollinear spin orders often show up in geometrically frustrated antiferromagnets as a compromise in minimizing the magnetic exchange energy. Such spin orders cause nontrivial physical phenomena~\cite{Nagaosa12}, e.g., the large anomalous Hall effect in the Mo pyrochlores with an umbrella-type spin order~\cite{Taguchi01,Taguchi03} and the magnetoelectric phenomena in the multiferroic Mn perovskites with a cycloidal spin order~\cite{Kimura03a,Katsura05,Cheong07,Tokura07}.
From intensive studies on these issues, we learn that microscopic spin models and a deep understanding of magnetism are crucially important in clarifying the physics behind the phenomena.

The inverse-perovskite structure $M_3AX$ [Fig.~\ref{Fig1}(a)] is one important example of a geometrically frustrated lattice. This crystal structure is a corner-sharing cubic network of the octahedra composed of six $M$ ions (transition metal). Each of the $X$ ions (light elements, e.g., H, B, C, N, O) is located at the center of an octahedron, whereas each of the $A$ ions (metal or semiconducting elements, e.g., Cu, Zn, Ga, Ge) is surrounded by eight $M$-octahedra. Because this crystal structure is basically composed of triangles of $M$ ions, antiferromagnetically interacting spins on this crystal lattice encounter significant frustration effects~\cite{Fruchart78,Kaneko87,Tahara07} and thereby can be a source of rich magnetism-driven phenomena such as the magnetovolume effect~\cite{Bouchaud68,Fruchart71,Takenaka12,Takenaka14}, large magnetostriction~\cite{Asano08,Takenaka10,WenYC10}, negative magnetocaloric effect~\cite{Tohei03,WangBS09}, and enhanced magnetoresistance~\cite{Kashima00,WangBS09b}.

Inverse-perovskite manganese nitrides Mn$_3$$A$N with $A$=Zn and Ga exhibit a dramatic negative thermal expansion~\cite{Bouchaud68,Fruchart71,Takenaka12,Takenaka14,Takenaka05,Takenaka06,Takenaka08,Hamada11,HuangR08,SunZH09,SunY07,SunY10,SongXY11}, that is, their crystal volume expands (shrinks) upon cooling (heating) in contrast to usual materials, which expand (shrink) as temperature increases (decreases). In 1978, it was experimentally uncovered that this sudden and pronounced increase in volume occurs when the material enters a triangular antiferromagnetic phase, labeled $\Gamma^{5g}$, from the paramagnetic phase through a magnetic phase transition upon cooling~\cite{Fruchart78}, although its origin have been unclarified almost for forty years. The spin configuration of this antiferromagnetic phase is shown in Fig.~\ref{Fig1}(b)~\cite{Iikubo08,Kodama10}. In addition to this $\Gamma^{5g}$ spin order, Mn$_3$$A$N also exhibits other types of magnetic order depending on the $A$-site species, specifically, a coexisting triangular $\Gamma^{4g}$ antiferromagnetic order in Mn$_3$NiN and Mn$_3$AgN [Fig.~\ref{Fig1}(c)]~\cite{Fruchart78} and a ferromagnetic order in Mn$_3$CuN~\cite{Takenaka14}.

To clarify the physics behind the observed unconventional magnetovolume effect in Mn$_3$ZnN and Mn$_3$GaN, the microscopic modeling of the spins as well as understanding the underlying magnetic behavior are essential. However, the origin of the variety of magnetic orders and a mechanism that stabilizes the triangular $\Gamma^{5g}$ spin order have remained as issues to be clarified since 1978. In addition, superconductivity has been discovered recently in inverse-perovskite nickelates Ni$_3$MgC~\cite{HeT01}, Ni$_3$CdC~\cite{Uehara06} and Ni$_3$ZnN$_y$~\cite{Uehara09}. Knowledge of the magnetism in the inverse-perovskite magnets may be useful also for understanding superconductivity because they are often closely related.

In this Letter, we construct a microscopic spin model for the inverse-perovskite manganese nitrides Mn$_3$$A$N by taking into account the frustrated exchange interactions and magnetic anisotropy specific to this class of materials. We argue that the introduced magnetic anisotropy is naturally expected for Mn$_3$$A$N from a consideration of electronic structures governed by its crystal symmetry. Numerical analyses of this spin model using the replica-exchange Monte-Carlo technique successfully reproduce the series of observed magnetic orders and the reported systematic evolution of the magnetic orders in Mn$_3$$A$N obtained experimentally. We also uncover the crucial role of the magnetic anisotropy in stabilizing the $\Gamma^{5g}$ and $\Gamma^{4g}$ spin orders. Our model and findings solve the 40-year-old problem of the nontrivial magnetic orders in the manganese inverse perovskites and provide a good starting point for research on the negative thermal expansion observed in this class of materials.

\begin{figure}[b]
\includegraphics[scale=0.9]{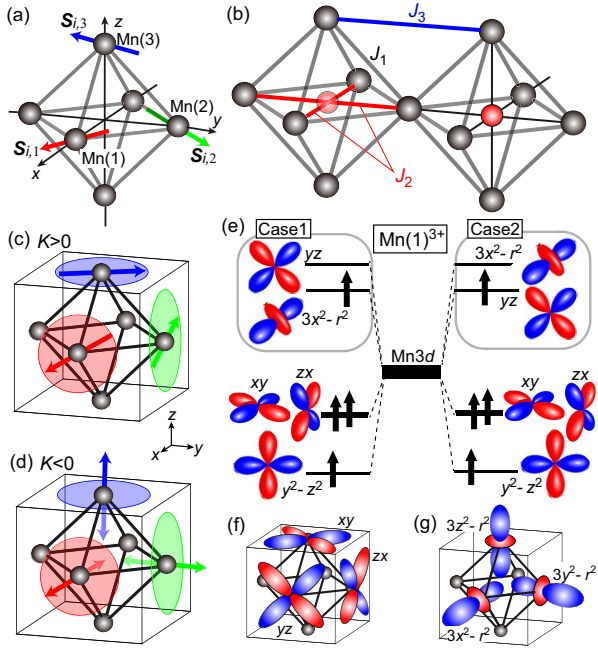}
\caption{(color online). (a) Three Mn sublattices Mn($\mu$) with spins $\bm S_{i,\mu}$ ($\mu$=1,2,3) on the $i$th octahedron. (b) Exchange interactions considered for the spin model~(\ref{eq:model}). (c) Easy-plane magnetic anisotropy with $K>0$. (d) Easy-axis magnetic anisotropy with $K<0$. (e) Orbital-level schemes of the Mn$^{3+}$ ion on the Mn(1) sublattice. Two cases, labeled 1 and 2, are possible depending on the ratio of the crystal field strengths for $A$ ions and for N ions acting on the Mn$3d$ orbitals; Case 1 (Case 2) obtains when the crystal field for $A$ (N) ions is stronger. (f)[(g)] Unoccupied orbitals of the highest levels on the three Mn sublattices for Case 1 [Case 2].}
\label{Fig2}
\end{figure}
Each unit cell of the inverse-perovskite lattice of Mn$_3$$A$N contains three different Mn sublattices Mn($\mu$) with $\mu$=1, 2 and 3 [Fig.~\ref{Fig2}(a)]. Both the triangular $\Gamma^{5g}$ and $\Gamma^{4g}$ spin orders are three-sublattice orders and their spin structures are easily visualized by considering a square cube, each face of which has a Mn ion at the center [Fig.~\ref{Fig1}(b) and (c)]. The spin vectors in the $\Gamma^{5g}$ pattern are lying in each face pointing along one of its diagonals. In contrast, the spin vectors in the $\Gamma^{4g}$ pattern are pointing towards the center of mass of the equilateral triangle formed by the diagonals of three adjoining faces, and therefore have out-of-face components. The sum of the three sublattice spin vectors (red, green, and blue arrows) vanishes for both patterns.

The physical properties of Mn$_3$$A$N are governed by the electronic structure near the Fermi level, which consists of a broad Mn$4s$ band and a narrow Mn$3d$--N$2p$ covalent band~\cite{Jardin83,Motizuki88}. The localized Mn$3d$ spins are mutually coupled via exchange interactions and therefore can be described by a classical Heisenberg model, whereas the itinerant Mn$4s$ conduction electrons move under the influence of potentials from a background Mn$3d$ spin texture mediated by the $s$--$d$ coupling. 

Spin-ordering patterns are strongly degenerate on the frustrated lattices. To reproduce the observed three-sublattice spin patterns by lifting the degeneracy, spins on the equivalent Mn sites must be parallel, and therefore ferromagnetic interactions are required for the next-nearest-neighbor bonds represented by $J_2$ and $J_3$ in Fig.~\ref{Fig2}(b). Note that the $J_2$ bond and the $J_3$ bond have the same length, but are inequivalent because the $J_2$ bond is mediated by a $X$(=N) ion, whereas the $J_3$ bond is not. In contrast, the nearest-neighbor coupling $J_1$ can be either antiferromagnetic or ferromagnetic. The sign of $J_1$ is governed by the $A$-site species via the orbital degeneracy to be explained below.

In the subspace of three-sublattice orders, all the $\mu$-th sublattice spins are equivalent by definition for $\mu$=1,2 and 3, and thus can be represented by a unified symbol $\bm S_\mu$ where the index of unit cells $i$ is eliminated. In this case, an energy contribution from the nearest-neighbor coupling $J_1$ can be written as
\begin{eqnarray}
& &4NJ_1 (\bm S_1 \cdot \bm S_2 + \bm S_2 \cdot \bm S_3 +\bm S_3 \cdot \bm S_1)
\notag\\
&=&2NJ_1 (\bm S_1 + \bm S_2 + \bm S_3)^2 + {\rm const},
\end{eqnarray}
because the numbers of nearest-neighbor sublattice pairs of ($\bm S_1$, $\bm S_2$), ($\bm S_2$, $\bm S_3$) and ($\bm S_3$, $\bm S_1$) in the whole system are all 2N, respectively. This formula indicates that the sum of the three sublattice spins, $\bm S_1 + \bm S_2 + \bm S_3$, for the lowest energy state is zero when $J_1>0$ (antiferromagnetic), whereas the spins are all parallel when $J_1<0$ (ferromagnetic). Both the $\Gamma^{5g}$ and $\Gamma^{4g}$ spin patterns satisfy the condition for $J_1>0$. However, the combinations of $\bm S_1$, $\bm S_2$ and $\bm S_3$ satisfying the condition $\bm S_1 + \bm S_2 + \bm S_3=0$ are all degenerate. Hence the spin ordering pattern in Mn$_3$$A$N cannot be determined by the $J_1$ coupling only.

To lift this degeneracy and reproduce the experimentally observed spin patterns, we introduce a magnetic anisotropy represented by,
\begin{eqnarray}
K \sum_{i, \mu} (\bm S_{i,\mu} \cdot \bm e_\mu)^2.
\end{eqnarray}
Here $\bm S_{i,\mu}$ denotes a classical spin vector on the $\mu$th Mn sublattice Mn($\mu$) in the $i$th octahedron. The norm of $\bm S_{i,\mu}$ is set to unity ($|\bm S_{i,\mu}|=1$). This term with $K>0$ [$K<0$] gives a hard [easy] magnetization axis parallel to a unit directional vector $\bm e_\mu$ on the Mn($\mu$) sublattice; see Fig.~\ref{Fig2}(c) [Fig.~\ref{Fig2}(d)]. The vector $\bm e_\mu$ differs depending on the sublattice; that is, the $\bm e_\mu$ vectors are $\bm \hat{x}$, $\bm \hat{y}$ and $\bm \hat{z}$ for Mn(1), Mn(2), and Mn(3) sites, respectively. 

\begin{figure}[t]
\includegraphics[scale=1.0]{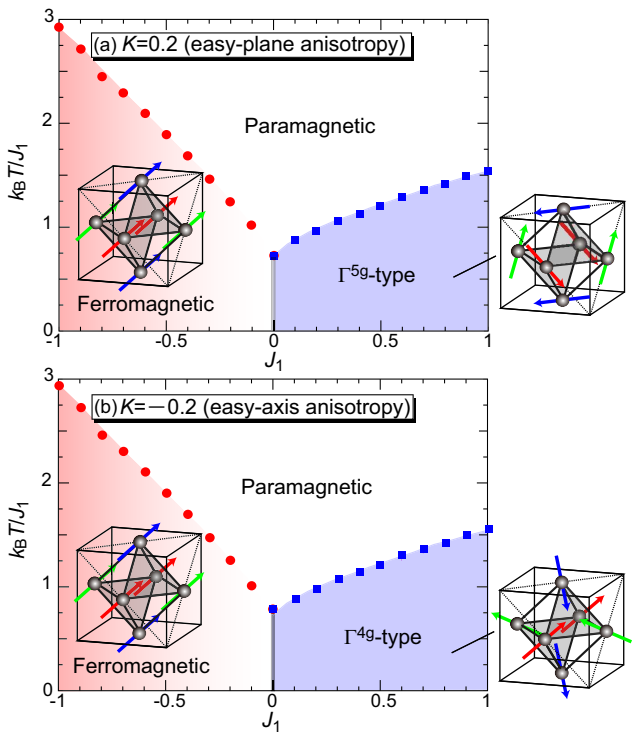}
\caption{(color online). Magnetic phase diagrams of the spin model~(\ref{eq:model}) in plane of $T$ and $J_1$ for (a) $K=0.2$ (easy-plane anisotropy) and (b) $K=-0.2$ (easy-axis anisotropy). 
}
\label{Fig3}
\end{figure}
We expect that the sign of $K$ varies depending on the $A$-site species. This sign variation as well as the emergence of this specific type of magnetic anisotropy in Mn$_3$$A$N can be understood by considering the energy-level schemes of the Mn$3d$ orbitals. From the crystallographic symmetry, the five-fold Mn$3d$ level splits into four levels [see Fig.~\ref{Fig2}(e)] where the second-lowest level is two-fold degenerate, whereas the other three levels have no degeneracy. Because the Mn$^{3+}$ ion has four $3d$ electrons and Hund's-rule coupling favors a high-spin state, the $3d$ orbitals up to the third level are almost occupied, whereas the highest (fourth) level is sparsely occupied. Furthermore, the orbital character of each level differs among the three Mn sublattices. Figure~\ref{Fig2}(e) shows two possible cases for the orbital character on the Mn(1) sublattice. We find that the realtive energy level of the $yz$ orbital pointing to the four $A$ ions and that of the $3x^2-r^2$ orbital pointing to the two N ions differ between Cases 1 and 2. Note that these two orbitals on the Mn(1) site become higher in energy due to the presence of the crystal field of the $A$ ions and that of the N ions, respectively. Competition between these two crystal fields governs the energy level relationship. When the crystal field of the $A$ (N) ions is stronger, the $yz$ ($3x^2-r^2$) orbital becomes higher in energy as in Case 1 (Case 2). In Case 1, the highest unoccupied orbitals on the Mn(1), Mn(2), and Mn(3) sublattices are $zx$, $yz$, and $xy$ orbitals, respectively [see Fig.~\ref{Fig2}(f)]. The spin-orbit couplings in these orbitals favor spins lying in the $zx$, $yz$, and $xy$ planes, respectively, and thus cause an easy-plane magnetic anisotropy; see Fig.~\ref{Fig2}(c). In Case 2, the highest unoccupied orbitals are $3x^2-r^2$, $3y^2-r^2$, and $3z^2-r^2$ orbitals for the Mn(1), Mn(2), and Mn(3) sublattices, respectively [see Fig.~\ref{Fig2}(g)], which produce an easy-axis magnetic anisotropy; see Fig.~\ref{Fig2}(d).

Based on the above consideration, we construct a classical Heisenberg model to describe the magnetism in Mn$_3$$A$N. The Hamiltonian is give by,
\begin{eqnarray}
\mathcal{H}=\sum_{i,\mu,j,\nu}J_{i\mu,j\nu} \bm S_{i,\mu} \cdot \bm S_{j,\nu} 
+ K \sum_{i, \mu} (\bm S_{i,\mu} \cdot \bm e_\mu)^2.
\label{eq:model}
\end{eqnarray}
The exchange-coupling coefficients $J_{i\mu,j\nu}$ are $J_1$ for the nearest-neighbor bonds, whereas they are $J_2(<0)$ [$J_3(<0)$] for the next-nearest-neighbor ferromagnetic bonds within the octahedron [between adjacent octahedra]. The information of the actual spin length is renormalized in the coefficients.

The nearest-neighbor coupling $J_1$ can be either antiferromagnetic or ferromagnetic depending on the $A$-site species. If the crystal field of the $A$ ions is stronger or weaker than that of the N ions, the energy splitting between the third and fourth levels becomes finite, resulting in the absence of orbital degrees of freedom. This gives rise to an antiferromagnetic coupling for the $J_1$ bonds ($J_1>0$). Moreover, if these two levels are degenerate with a subtle balance between the two crystal fields, the $J_1$ coupling should be ferromagnetic because Hund's--rule coupling favors the ferromagnetic coupling in the presence of the orbital degeneracy.

The above spin model is analyzed using a replica-exchange Monte-Carlo method. For the calculations, we adopt systems of $3L^3$ spin sites with a periodic boundary condition where $L^3$ is the number of Mn$_6$N octahedra. The next-nearest-neighbor ferromagnetic couplings $J_2$ and $J_3$ are fixed at $J_2=J_3=-0.5$.
\begin{figure}
\includegraphics[scale=1.0]{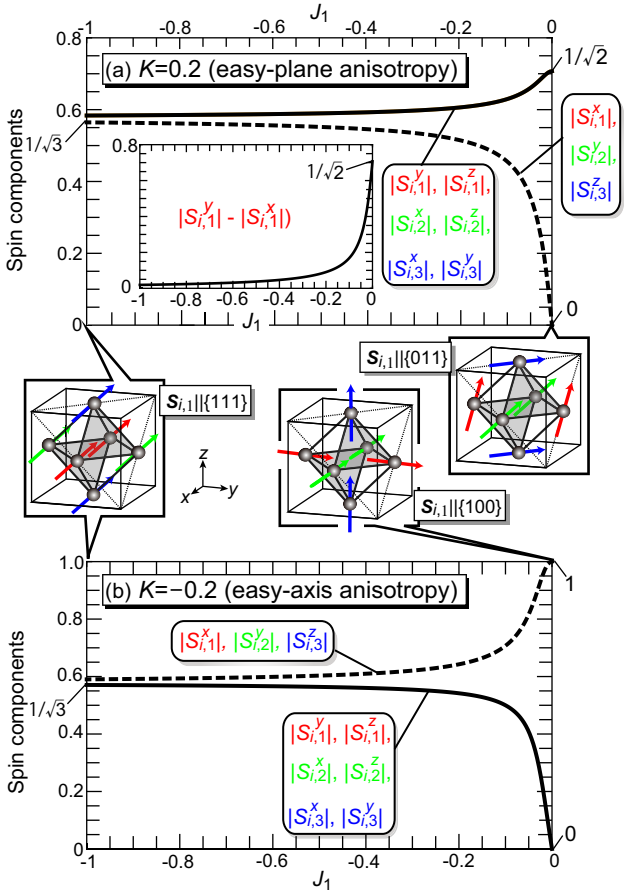}
\caption{(color online). (a) Components of the spin vectors $\bm S_{i,\mu}$, i.e., $|S_{i,\mu}^x|$,$|S_{i,\mu}^y|$,$|S_{i,\mu}^z|$ on the Mn($\mu$) sublattice at $T$=0, as functions of $J_1(<0)$ for easy-plane magnetic anisotropy with $K=0.2$. (b) Those for easy-axis magnetic anisotropy with $K=-0.2$. The spin ordering patterns for the weak ferromagnetic-coupling limit ($|J_1/K| \ll 1$) and those for the strong ferromagnetic-coupling limit ($|J_1/K| \gg 1$) are also displayed.}
\label{Fig4}
\end{figure}
In Fig.~\ref{Fig3}(a), we display a magnetic phase diagram constructed for temperature ($k_{\rm B}T/J_1$) and coupling $J_1$ with $K=0.2$ (easy-plane anisotropy) for a system size of $L$=12. The triangular $\Gamma^{5g}$ spin phase, which has been observed experimentally in Mn$_3$ZnN and Mn$_3$GaN, indeed takes place when $J_1>0$, whereas the ferromagnetic order is obtained if $J_1<0$. A phase boundary between these two phases is located exactly at $J_1$=0. Comparison with the experimentally reported magnetic transition temperatures of 100-300 K~\cite{MWu13} for the materials with $\Gamma^{5g}$ spin order, the values of $J_1$ in these materials are evaluated to be 10-30 meV.

On the other hand, the magnetic phase diagram for $K=-0.2$ (easy-axis anisotropy) [Fig.~\ref{Fig3}(b)] exhibits a magnetic phase transition from the triangular $\Gamma^{4g}$ spin phase to a ferromagnetic phase with decreasing $J_1$ from positive to negative. Their phase boundary is again located at $J_1$=0. 

It should be mentioned that the spin-ordering patterns in the ferromagnetic phases are not straightforward. Specifically, the orientations of the three sublattice spins continuously vary as $J_1(<0)$ decreases (equivalently, as its absolute value $|J_1|$ increases), reflecting a competition between the ferromagnetic coupling $J_1$ and the magnetic anisotropy $K$. Figure~\ref{Fig4}(a) gives $x$-, $y$-, and $z$-axis components of the spin vector $\bm S_{i,\mu}$ for each Mn subalttice ($\mu$=1,2,3) at $T$=0 as functions of $J_1$ when the magnetic anisotropy is the easy-plane type with $K>0$. We find that the spin vectors are lying within each face of the cubic unit cell if the ferromagnetic coupling $J_1$ is sufficiently weak that the easy-plane anisotropy dominates ($|J_1/K| \ll 1$). In turn, they all point nearly to the $\{111\}$-direction when the ferromagnetic coupling $J_1$ dominates ($|J_1/K| \gg 1$). From the difference between these two plots [inset of Fig~\ref{Fig4}(a)], the spin vectors rapidly polarize along the trigonal direction or the $\{111\}$ direction as $|J_1|$ increases. In contrast, the spin components for $K<0$ (easy-axis anisotropy) [Fig.~\ref{Fig4}(b)] indicate that the spin vectors are perpendicular to each face of the cube if $|J_1/K| \ll 1$, whereas the spin vectors are again polarized along the $\{111\}$ direction if $|J_1/K| \gg 1$. We mention that this kind of cross-over behavior can also be observed in the $\Gamma^{4g}$ phase for $K<0$.

\begin{figure}
\includegraphics[scale=1.0]{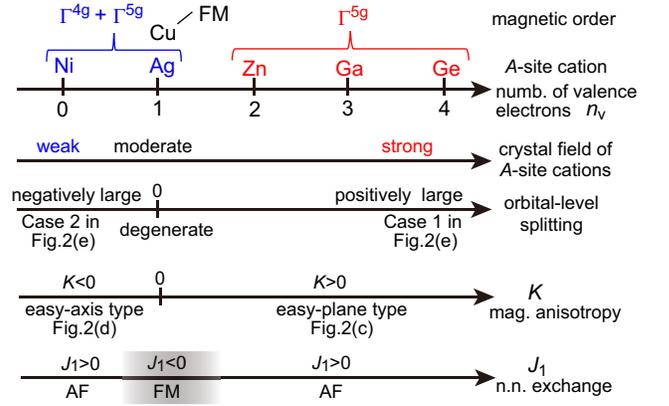}
\caption{(color online). Systematic evolution of the magnetic orders upon variation of the $A$-site species or the number of valence electrons. The $A$-site ions govern the signs of the magnetic anisotropy $K$ and the nearest-neighbor coupling $J_1$ by modulating the orbital-level schemes via generating the crystal field, which competes with the crystal field of the N ions.}
\label{Fig5}
\end{figure}
We now discuss the systematic evolution of the magnetic orders experimentally observed in Mn$_3$$A$N for various $A$-site cations. In the early stage of the research, Fruchart and Bertaut claim that there is a close relationship between the magnetic properties (e.g., the magnetic transition temperatures and the volume magnetostriction) and the number of valence electrons $n_{\rm v}$ in the $A$ ion. We summarize the $n_{\rm v}$-dependence of the magnetic-ordering patterns in Fig.~\ref{Fig5}. Note that the $\Gamma^{5g}$ ($\Gamma^{4g}$) order tends to appear when $n_{\rm v}$ is large (small), whereas in between ferromagnetic order obtains. This tendency is understandable if we assume that the magnitude of the $A$-ion crystal field becomes stronger as $n_{\rm v}$ is larger because of stronger repulsive Coulomb potentials from the valence electrons. When the crystal field of the $A$ ions is stronger (weaker) with a larger (smaller) $n_{\rm v}$, the orbital-level scheme of Case 1 (Case 2) in Fig.~\ref{Fig2}(e) is realized, which results in a magnetic anisotropy of the easy-plane type (easy-axis type) with $K>0$ ($K<0$); see Fig.~\ref{Fig2}(c)[(d)]. In addition, the nearest-neighbor coupling $J_1$ should be antiferromagnetic if the orbitals are non-degenerate with unbalanced crystal fields from the $A$ and $N$ ions. Consequently, the magnetic order tends to be pure $\Gamma^{5g}$-type for $A$=Zn ($n_{\rm v}$=2) and Ga ($n_{\rm v}$=3), whereas a mixture of the $\Gamma^{4g}$-type occurs for $A$=Ni ($n_{\rm v}$=0) and Ag ($n_{\rm v}$=1). When the crystal field of the $A$ ions is moderate in strength and is comparable to the crystal field of the $N$ ions, the third and fourth orbital levels become nearly degenerate, which induces a negligibly weak magnetic anisotropy ($K$$\sim$0) and a ferromagnetic $J_1$ coupling ($J_1<0$), resulting in ferromagnetic order for $A$=Cu ($n_{\rm v}$=1). This argument also indicates that $J_1$ coupling and magnetic anisotropy $K$ are not independent of each other but are closely related via the electronic structure governed by the two competing crystal fields.

In summary, a classical spin model with frustrated exchange interactions and magnetic anisotropy was constructed to study the nontrivial magnetic orders in the inverse-perovskite manganese nitrides Mn$_3$$A$N taking into account the electronic structure in this specific crystal lattice. Analyzing this spin model using Monte-Carlo methods, the experimentally observed triangular $\Gamma^{5g}$ and $\Gamma^{4g}$ spin ordering patterns have been reproduced, which are known to trigger the unusual magnetovolume effect, i.e., negative thermal expansion. To fully clarify this magnetism-driven volume expansion phenomenon, we need to further take into account coupling between the magnetism and lattice degrees of freedom. The present work will provide a firm basis for future research in this direction.

This research was supported by JSPS KAKENHI (Grant Nos. 25870169, 25287088, 26800200, and 17H02924), Waseda University Grant for Special Research Projects (Project No. 2017S-101), and JST PRESTO (Grant No. JPMJPR132A). 

\end{document}